\documentclass[aps,showpacs]{revtex4}
\usepackage{graphicx}
\begin{document}

\title{Centrality dependence of global variables in relativistic heavy ion collisions:\\
Final $p_{T}$ data analysis in the framework of a statistical
model}
\author{Dariusz Prorok}
\email{prorok@ift.uni.wroc.pl} \affiliation{Institute of
Theoretical Physics, University of Wroc{\l}aw,\\ Pl.Maksa Borna 9,
50-204 Wroc{\l}aw, Poland}
\date{April 19, 2006}

\begin{abstract}
The global variables like the transverse energy at midrapidity,
the charged particle multiplicity at midrapidity and the total
multiplicity of charged particles are evaluated in the
single-freeze-out statistical model for different centrality bins
at RHIC at $\sqrt{s_{NN}}=130$ and 200 GeV. Full description of
decays of hadron resonances is applied in these estimations. The
geometric parameters of the model are obtained from the fit to the
final data on the $p_{T}$ spectra. The predicted values of the
global variables agree qualitatively well with the experimental
data. The centrality independence of the total number of charged
particles per participant pair has been also reproduced.
\end{abstract}

 \maketitle

\section {Introduction}

In the previous paper \cite{Prorok:2004wi} the extensive analysis
of the centrality dependence of two measured global variables,
transverse energy ($dE_{T}/d\eta\vert_{mid}$) and charged particle
multiplicity ($dN_{ch}/d\eta\vert_{mid}$) densities at
mid-rapidity as well as their ratio, was delivered. The analysis
was done in the framework of the single freeze-out statistical
model \cite{Florkowski:2001fp,Broniowski:2001we,Broniowski:2001uk}
for the Au-Au collisions at RHIC at $\sqrt{s_{NN}}=130$ and
200~GeV. The main idea of this method is as follows. Thermal and
geometric parameters of the model are established from fits to
particle yield ratios and $p_{T}$ spectra, respectively. Then,
with the use of these parameters both densities $dE_{T}/d\eta$ and
$dN_{ch}/d\eta$ as well as the total charged-particle multiplicity
can be estimated and compared with the data. The main reason for
doing it is that the transverse energy and charged particle
multiplicity measurements are independent of hadron spectroscopy
(in particular, no particle identification is necessary),
therefore they could be used as an additional test of the
self-consistency of a statistical model. In the present paper the
above-mentioned program will be performed once more for the
reasons stated below.

First of all fits of the geometric parameters \cite{Baran:2003nm}
were done with the use of the preliminary data for the $p_{T}$
spectra measured at RHIC at $\sqrt{s_{NN}}=200$ GeV
\cite{Chujo:2002bi,Barannikova:2002qw}. But the final data
\cite{Adler:2003cb,Adams:2003xp} differ substantially from the
preliminary ones. Additionally the data points for the $\chi^{2}$
analysis were digitized from the plots. Also not all centrality
bins were fitted in \cite{Baran:2003nm,Broniowski:2002nf}. All of
these resulted in the rough qualitative agreement of the model
predictions for $dN_{ch}/d\eta$ and $dE_{T}/d\eta$ for RHIC at
$\sqrt{s_{NN}}=200$ GeV \cite{Prorok:2004wi}. And the centrality
independence of the total multiplicity of charged particles per
participant pair was not reproduced. The evidence for such scaling
of the total multiplicity was reported by the PHOBOS Collaboration
\cite{Back:2003xk}. Therefore it was very tempting to check
whether the $\chi^{2}$ analysis of the final data for the $p_{T}$
spectra of identified charged hadrons would improve the
predictions for transverse energy and charged particle
multiplicity densities and the total charged-particle
multiplicity. Such a check is \textit{nontrivial} and \textit{does
not mean} simply adding the integrals of the experimental momentum
distributions because: (\textit{a}) spectra are measured in the
limited ranges of $p_{T}$ and the very important low-$p_{T}$ part
is not detected, (\textit{b}) spectra of stable neutral particles
like neutrons and $K_{L}^{0}$ are not measured generally.

As a statistical model the single freeze-out model is taken (for
details see \cite{Broniowski:2002nf}). The model succeeded in the
accurate description of ratios and $p_{T}$ spectra of particles
measured at RHIC
\cite{Florkowski:2001fp,Broniowski:2001we,Broniowski:2001uk}. The
main postulate of the model is the simultaneous occurrence of
chemical and thermal freeze-outs, which means that the possible
elastic interactions after the chemical freeze-out are neglected.
The conditions for the freeze-out are expressed by values of two
independent thermal parameters: $T$ and $\mu_{B}$. The strangeness
chemical potential $\mu_{S}$ is determined from the requirement
that the overall strangeness equals zero. The second basic feature
of the model is the complete treatment of resonance decays. This
means that the final distribution of a given particle consists not
only of the thermal part but also of contributions from all
possible decays and cascades. Feeding from week decays is included
as well.

\section { The foundations of the single-freeze-out model }
\label{Foundat}

The main assumptions of the model are as follows: (\textit{a}) the
chemical and thermal freeze-outs take place simultaneously,
(\textit{b}) all confirmed resonances up to a mass of $2$ GeV from
the Particle Data Tables \cite{Hagiwara:fs} are taken into
account, (\textit{c}) a freeze-out hypersurface is defined by the
equation

\begin{equation}
\tau = \sqrt{t^{2}-r_{x}^{2}-r_{y}^{2}-r_{z}^{2}}= const \;,
\label{Hypsur}
\end{equation}

\noindent (\textit{d}) the four-velocity of an element of the
freeze-out hypersurface is proportional to its coordinate

\begin{equation}
u^{\mu}={ {x^{\mu}} \over \tau}= {t \over \tau}\; \left(1,{
{r_{x}} \over t},{{r_{y}} \over t},{{r_{z}} \over t}\right) \;,
\label{Velochyp}
\end{equation}

\noindent (\textit{e}) the following parameterization of the
hypersurface is chosen:

\begin{equation}
t= \tau \cosh{\alpha_{\parallel}}\cosh{\alpha_{\perp}},\;\;\;
r_{x}=  \tau \sinh{\alpha_{\perp}}\cos{\phi},\;\;\; r_{y}=  \tau
\sinh{\alpha_{\perp}}\sin{\phi},\;\;\;r_{z}=\tau
\sinh{\alpha_{\parallel}}\cosh{\alpha_{\perp}}, \label{Parahyp}
\end{equation}

\noindent where $\alpha_{\parallel}$ is the rapidity of the
element, $\alpha_{\parallel}= \tanh^{-1}(r_{z}/t)$, and
$\alpha_{\perp}$ controls the transverse radius:

\begin{equation}
\rho= \sqrt{r_{x}^{2}+r_{y}^{2}}= \tau \sinh{\alpha_{\perp}} <
\rho_{max} \;, \label{Transsiz}
\end{equation}

\noindent where the restriction on the transverse size has been
introduced, so $\rho_{max}$ gives the maximal transverse extension
of the gas in the central slice during the freeze-out. This means
that two new parameters of the model have been introduced,
\emph{i.e.} $\tau$ and $\rho_{max}$, which are connected with the
geometry of the freeze-out hypersurface. In general a freeze-out
addresses a 6-dimensional phase-space but the assumption of the
Hubble-like expansion (the velocity at the freeze-out is
proportional to the coordinate), Eq.~(\ref{Velochyp}), reduces the
number of independent variables to 3. And the assumption of the
cylindrical symmetry further reduces this number to 2. Another
point is how realistic the assumption of the Hubble flow is.
Recent results of Refs.~\cite{Csanad:2004mm,Chojnacki:2004ec}
indicate that the Hubble flow of the form of Eq.~(\ref{Velochyp})
can develop in 130 and 200 GeV Au-Au collisions within times 6-15
fm. This is roughly the scale of the freeze-out initializing time
predicted in this analysis for all but peripheral bins (see
Table~\ref{Table1}).

From Eq.~(\ref{Hypsur}) one can see that the beginning of the
freeze-out process starts at $t_{f.o.}^{(1)}=\tau$ and $\vec{r}=0$
in the c.m.s., which is also the laboratory frame in the RHIC
case. At this moment the volume of the gas can be estimated as

\begin{equation}
V_{f.o.}^{(1)} = 2\pi\tau\rho_{max}^{2} \;, \label{Volgas}
\end{equation}

\noindent which is simply the volume of a tube with a length
$2\tau$ and a radius $\rho_{max}$ ($2\tau$ is the maximal possible
extension of the gas in the longitudinal direction at
$t_{f.o.}^{(1)}$). In the central slice the freeze-out ceases at
$t_{f.o.}^{(2)}= \sqrt{\tau^{2}+\rho_{max}^{2}}$ and it takes
place at $\rho=\rho_{max}$. This \textit{does not mean} the end of
the freeze-out process completely. It is the moment when the gas
decouples into two pieces symmetrical with respect to the plane
$z=0$ and from then the freeze-out proceeds into two opposite
directions of the collision axis $z$. For the considered
hypersurface, Eq.~(\ref{Hypsur}), this process extends to infinity
but to be more realistic a limiting condition for
$\alpha_{\parallel}$ will be assumed in Sec.~\ref{Totalcharge},
namely a maximal possible value of $\alpha_{\parallel}$ called
$\alpha_{\parallel}^{max}$ will be postulated. Then the moment
when the freeze-out ceases at all will be equal to

\begin{equation}
t_{f.o.}^{(3)} = \cosh{\alpha_{\parallel}^{max}} \cdot
t_{f.o.}^{(2)} \;. \label{Timefofin}
\end{equation}

\noindent At $t_{f.o.}^{(2)}$ the volume of the region where the
freeze-out process has just taken place or is to happen can be
estimated as

\begin{equation}
V_{f.o.}^{(2)} = 2\pi \sqrt{\tau^{2}+\rho_{max}^{2}}\;
\rho_{max}^{2} \;. \label{Volgas2}
\end{equation}

The transverse velocity in the central slice can be expressed as a
function of the transverse radius

\begin{equation}
\beta_{\perp}(\rho)= \tanh{\alpha_{\perp}}= { \rho \over
{\sqrt{\tau^{2}+\rho^{2}}}}\;. \label{Betprof1}
\end{equation}

\noindent The maximum value of $\beta_{\perp}$ called the maximum
transverse-flow parameter (or the surface velocity) is given by

\begin{equation}
\beta_{\perp}^{max}= { \rho_{max} \over
{\sqrt{\tau^{2}+\rho_{max}^{2}}}}= { {\rho_{max}/\tau} \over
{\sqrt{1+(\rho_{max}/\tau)^{2}}}}\;. \label{Betmax}
\end{equation}

The invariant distribution of the measured particles of species
$i$ has the form \cite{Broniowski:2001we,Broniowski:2001uk}

\begin{equation}
{ {dN_{i}} \over {d^{2}p_{T}\;dy} }=\int
p^{\mu}d\sigma_{\mu}\;f_{i}(p \cdot u) \;, \label{Cooper}
\end{equation}

\noindent where $d\sigma_{\mu}$ is the normal vector on a
freeze-out hypersurface, $p \cdot u = p^{\mu}u_{\mu}$ , $u_{\mu}$
is the four-velocity of a fluid element and $f_{i}$ is the final
momentum distribution of the particle in question. The final
distribution means here that $f_{i}$ is the sum of primordial and
simple and sequential decay contributions to the particle
distribution (for details see
\cite{Prorok:2004af,Broniowski:2002nf}).

With the use of Eqs.~(\ref{Velochyp}) and (\ref{Parahyp}), the
invariant distribution (\ref{Cooper}) takes the following form:

\begin{equation}
{ {dN_{i}} \over {d^{2}p_{T}\;dy} }= \tau^{3}\;
\int\limits_{-\infty}^{+\infty}
d\alpha_{\parallel}\;\int\limits_{0}^{\rho_{max}/\tau}\;\sinh{\alpha_{\perp}}
d(\sinh{\alpha_{\perp}})\; \int\limits_{0}^{2\pi} d\xi\;(p \cdot
u) \; f_{i}(p \cdot u) \;, \label{Cooper2}
\end{equation}

\noindent where

\begin{equation}
p \cdot u =
m_{T}\cosh{(\alpha_{\parallel}-y)}\cosh{\alpha_{\perp}}-
p_{T}\cos{\xi}\sinh{\alpha_{\perp}}\;. \label{Peu}
\end{equation}

\noindent After changing the integration variable
$\alpha_{\parallel}$ to $\alpha_{\parallel}+y$ the invariant
distribution (\ref{Cooper2}) still has the same form but with $p
\cdot u$ replaced now by $\widetilde{p \cdot u}$:

\begin{equation}
\widetilde{p \cdot u} =
m_{T}\cosh{\alpha_{\parallel}}\cosh{\alpha_{\perp}}-
p_{T}\cos{\xi}\sinh{\alpha_{\perp}}\;. \label{Peu2}
\end{equation}

\noindent That was for the case of unlimited $\alpha_{\parallel}$.
If $\alpha_{\parallel}$ is limited and has its maximal value
$\alpha_{\parallel}^{max}$ then the invariant distribution
(\ref{Cooper}) reads

\begin{equation}
{ {dN_{i}} \over {d^{2}p_{T}\;dy} }= \tau^{3}\;
\int\limits_{-\alpha_{\parallel}^{max}}^{\alpha_{\parallel}^{max}}
d\alpha_{\parallel}\;\int\limits_{0}^{\rho_{max}/\tau}\;\sinh{\alpha_{\perp}}
d(\sinh{\alpha_{\perp}})\; \int\limits_{0}^{2\pi} d\xi\;(p \cdot
u) \; f_{i}(p \cdot u) = \int d\sigma\;(p \cdot u)\;f_{i}(p \cdot
u) \;. \label{Cooper3}
\end{equation}

\noindent The same changing of the integration variable
$\alpha_{\parallel} \rightarrow\alpha_{\parallel}+y$ can be done
and one obtains

\begin{equation}
{ {dN_{i}} \over {d^{2}p_{T}\;dy} }= \tau^{3}\;
\int\limits_{-\alpha_{\parallel}^{max}-y}^{\alpha_{\parallel}^{max}-y}
d\alpha_{\parallel}\;\int\limits_{0}^{\rho_{max}/\tau}\;\sinh{\alpha_{\perp}}
d(\sinh{\alpha_{\perp}})\; \int\limits_{0}^{2\pi}
d\xi\;(\widetilde{p \cdot u}) \; f_{i}(\widetilde{p \cdot u}) \;.
\label{Cooper4}
\end{equation}

\noindent The distribution (\ref{Cooper4}) is still boost
invariant because rapidity differences are invariant with respect
to the longitudinal boosts.

In practical, \textit{i.e.} numerical, calculations of the
invariant distribution given by Eqs.~(\ref{Cooper2}) and
(\ref{Peu2}) a cut in the integral variable $\alpha_{\parallel}$
has to be done. So for the numerical reasons the range of
$\alpha_{\parallel}$ is always finite and
$-\alpha_{\parallel}^{cut} \leq \alpha_{\parallel} \leq
\alpha_{\parallel}^{cut}$. The value of $\alpha_{\parallel}^{cut}$
is chosen in such a way that the remaining "tails" of the integral
over $\alpha_{\parallel}$ are negligible. It has turned out that
it is enough to put $\alpha_{\parallel}^{cut}=3.0$ to determine
the geometric parameters $\tau$ and $\rho_{max}$ exact to four
decimal places.

It should be reminded here that in this paper fits are done to the
spectra measured at midrapidity ($y=0$). Also the estimates for
$dE_{T}/d\eta$ and $dN_{ch}/d\eta$ are done for $y=0$. What is
more, the values of $\alpha_{\parallel}^{max}$ predicted in
Sec.~\ref{Totalcharge} are greater than 3.0. Thus for the
practical calculations at $y=0$, the case with unlimited
$\alpha_{\parallel}$ covers the case with the finite range of
$\alpha_{\parallel}$.

\section {Transverse energy and charged particle
multiplicity densities} \label{Etncheta}

The experimentally measured transverse energy is defined as

\begin{equation}
E_{T} = \sum_{i = 1}^{L} \hat{E}_{i} \cdot \sin{\theta_{i}} \;,
\label{Etdef}
\end{equation}

\noindent where $\theta_{i}$ is the polar angle, $\hat{E}_{i}$
denotes $E_{i}-m_{N}$ ($m_{N}$ means the nucleon mass) for
baryons, $E_{i}+m_{N}$ for antibaryons and the total energy
$E_{i}$ for all other particles, and the sum is taken over all $L$
emitted particles \cite{Adams:2004cb,Adler:2004zn}.

The pseudorapidity density of particle species $i$ is given by

\begin{equation}
{ {dN_{i}} \over {d\eta} } = \int d^{2}p_{T}\; {{dy} \over {d\eta}
} \; { {dN_{i}} \over {d^{2}p_{T}\;dy} }= \int d^{2}p_{T}\; {p
\over {E_{i}} } \; { {dN_{i}} \over {d^{2}p_{T}\;dy} }\;.
\label{Partdens}
\end{equation}

\noindent Analogously, the transverse energy pseudorapidity
density for the same species can be written as

\begin{equation}
{ {dE_{T,i}} \over {d\eta} } = \int d^{2}p_{T}\; \hat{E}_{i} \cdot
{{p_{T}} \over p} \; {{dy} \over {d\eta} }\; { {dN_{i}} \over
{d^{2}p_{T}\;dy} }= \int d^{2}p_{T}\;{p_{T}} \; { {\hat{E}_{i}}
\over {E_{i}} }\; { {dN_{i}} \over {d^{2}p_{T}\;dy} }\;.
\label{Etraden}
\end{equation}

\noindent For the quantities at midrapidity one has (in the
c.m.s., which is the RHIC case)

\begin{equation}
{ {dN_{i}} \over {d\eta} }\;\Big\vert_{mid}= \int d^{2}p_{T}\; {
{p_{T}} \over {m_{T}} } \;{ {dN_{i}} \over {d^{2}p_{T}\;dy} }\; ,
\label{Partdenmid}
\end{equation}

\begin{equation}
{ {dE_{T,i}} \over {d\eta} }\;\Big\vert_{mid} = \cases{ \int
d^{2}p_{T}\;{p_{T}} \; { {m_{T}-m_{N}} \over {m_{T}} }\; {
{dN_{i}} \over {d^{2}p_{T}\;dy} }\;, i=baryon
 \cr \cr \int
d^{2}p_{T}\;{p_{T}}\; { {m_{T}+m_{N}} \over {m_{T}} } \;{ {dN_{i}}
\over {d^{2}p_{T}\;dy} }\;, i=antibaryon
 \cr \cr \int d^{2}p_{T}\;{p_{T}} \;{ {dN_{i}} \over
{d^{2}p_{T}\;dy} }, i=others  \;.} \label{Etdenmid}
\end{equation}

The overall charged particle and transverse energy densities can
be expressed as

\begin{equation}
{ {dN_{ch}} \over {d\eta} }\;\Big\vert_{mid}= \sum_{i \in B} {
{dN_{i}} \over {d\eta} }\;\Big\vert_{mid}\;, \label{Nchall}
\end{equation}

\begin{equation}
{ {dE_{T}} \over {d\eta} }\;\Big\vert_{mid}= \sum_{i \in A} {
{dE_{T,i}} \over {d\eta} }\;\Big\vert_{mid} \;, \label{Etall}
\end{equation}

\noindent where $A$ and $B$ ($B \subset A$) denote sets of species
of finally detected particles, namely the set of charged particles
$B=\{\pi^{+},\; \pi^{-},\; K^{+},\; K^{-},\; p,\; \bar{p}\}$,
whereas $A$ also includes photons, $K_{L}^{0},\; n$ and
$\bar{n}\;$ \cite{Adcox:2001ry}.

\section {Results} \label{Finl}

\subsection {Determination of geometric parameters} \label{Finlrhotau}

Analyses of the particle ratios and $p_{T}$ spectra at various
centralities in the framework of the single freeze-out model were
done for the preliminary RHIC data at $\sqrt{s_{NN}}=200$ GeV
\cite{Chujo:2002bi,Barannikova:2002qw} in \cite{Baran:2003nm}.
This approach proceeds in two steps. First, thermal parameters $T$
and $\mu_{B}$ are fitted with the use of the experimental ratios
of hadron multiplicities at midrapidity. It is assumed that these
values are independent of the centrality. This is reasonable since
the very weak centrality dependence of the particle ratios has
been observed so far. Recent works of
Ref.~\cite{Cleymans:2004pp,Rafelski:2004dp,Barannikova:2005rw}
have just confirmed this assumption. After then two next
parameters, $\tau$ and $\rho_{max}$, are determined from the
simultaneous fit to the transverse-momentum spectra of
$\pi^{\pm}$, $K^{\pm}$, $p$ and $\bar{p}$. Both fits are performed
with the help of the $\chi^{2}$ method. Since the preliminary data
for the $p_{T}$ spectra \cite{Chujo:2002bi,Barannikova:2002qw}
differ from the final data \cite{Adler:2003cb,Adams:2003xp} and
not all bins were fitted in \cite{Baran:2003nm}, the fit procedure
for determination of the geometric parameters of the model, $\tau$
and $\rho_{max}$, has been performed again for the purposes of
this paper. Additionally, the PHENIX case at $\sqrt{s_{NN}}=130$
GeV has been worked out once more, since the first published data
were for three bins only \cite{Adcox:2001mf}. The data for the
next two bins, which were not fitted in \cite{Broniowski:2002nf},
were added in the later report \cite{Adcox:2003nr}. Also the
BRAHMS spectra at various centralities for the measurement at
$\sqrt{s_{NN}}=200$ GeV \cite{Arsene:2005mr} have not been fitted
within the single-freeze-out model until now. The data points with
$p_{T} > 3$ GeV have been excluded in the $\chi^{2}$ analysis. The
thermal parameters for the three cases of the maximal RHIC
collision energy have been taken from the newer studies of the
particle abundance ratios
\cite{Rafelski:2004dp,Barannikova:2005rw}. According to these
results, the thermal parameters are different for two experiments,
PHENIX and STAR, in spite of the same collision energy. Fits to
mostly PHENIX ratios gave the lower freeze-out temperature
\cite{Rafelski:2004dp} than fits to STAR data only
\cite{Barannikova:2005rw}. This observation has been confirmed in
the present analysis: fits to the PHENIX spectra with substitution
of the STAR freeze-out thermal parameters ($T = 160.0$ MeV and
$\mu_{B} = 24.0$ MeV \cite{Barannikova:2005rw}) and fits to the
STAR spectra with the use of the PHENIX freeze-out thermal
parameters ($T = 155.2$ MeV and $\mu_{B} = 26.4$ MeV
\cite{Rafelski:2004dp}) have proven to be much worse than the fits
done within the same experiment and finally reported in this work
(here "worse fit" means that its $\chi^{2}$/NDF is greater). Also
it has turned out that BRAHMS spectra fit better with the PHENIX
thermal parameters. The older estimates of thermal parameters for
RHIC at $\sqrt{s_{NN}}=200$ GeV ($T = 165.6$ MeV and $\mu_{B} =
28.5$ MeV) given in Ref.~\cite{Baran:2003nm} have proven to lead
to fits to PHENIX spectra which are not statistically significant
for all centralities, $\chi^{2}$/NDF $\geq 1.3$. It should be also
mentioned at this point, that the values of thermal parameters ($T
= 177$ MeV and $\mu_{B} = 29$ MeV) obtained in
Ref.~\cite{Braun-Munzinger:2003zd} result in fits which are much
worse, $\chi^{2}$/NDF $\approx 3$.

\begin{table}
\caption{\label{Table1} Values of the geometric parameters of the
model for various centrality bins fitted with the use of the RHIC
final data for the $p_{T}$ spectra of identified charged hadrons
\protect\cite{Adcox:2003nr,Adler:2003cb,Adams:2003xp,Arsene:2005mr,Bearden:2003hx}.
Values of the thermal parameters are taken from the quoted
references. }
\begin{ruledtabular}
\begin{tabular}{cccccccccc} \hline Collision case & Centrality [\%] &
$N_{part}$ & $\rho_{max}$ [fm] & $\tau$ [fm] &
$\beta_{\perp}^{max}$ & $V_{f.o.}^{(1)}$ [fm$^{3}$] &
$t_{f.o.}^{(2)}$ [fm] & $V_{f.o.}^{(2)}$ [fm$^{3}$] &
$\chi^{2}$/NDF
\\
\hline PHENIX at & 0-5 & 347.7 & 6.50$\pm$0.27 & 8.23$\pm$0.23 &
0.62 & 2184.1 & 10.5 & 2782.2 & 0.52
\\
$\sqrt{s_{NN}}=130$ GeV: & 5-15 & 271.3 & 5.99$\pm$0.21 &
7.29$\pm$0.18 & 0.63 & 1641.1 & 9.4 & 2124.2 & 0.46
\\
$T = 165$ MeV & 15-30 & 180.2 & 5.08$\pm$0.18 & 6.34$\pm$0.15 &
0.63 & 1028.1 & 8.1 & 1317.6 & 0.49
\\
$\mu_{B} = 41$ MeV  & 30-60 & 78.5 & 3.59$\pm$0.15 & 4.81$\pm$0.13
& 0.60 & 388.9 & 6.0 & 485.4 & 0.74
\\
\protect\cite{Florkowski:2001fp} & 60-92 & 14.3 & 1.68$\pm$0.19 &
3.14$\pm$0.22 & 0.47 & 55.6 & 3.6 & 63.1 & 1.32
\\
\hline PHENIX at & 0-5 & 351.4 & 8.46$\pm$0.10 & 8.84$\pm$0.08 &
0.69 & 3973.4 & 12.2 & 5497.0 & 0.80
\\
$\sqrt{s_{NN}}=200$ GeV: & 5-10 & 299.0 & 7.99$\pm$0.10 &
8.23$\pm$0.08 & 0.70 & 3302.6 & 11.5 & 4602.9 &  0.61
\\
$T = 155.2$ MeV & 10-15 & 253.9 & 7.54$\pm$0.10 & 7.67$\pm$0.08 &
0.70 & 2736.2 & 10.8 & 3837.1 & 0.48
\\
$\mu_{B} = 26.4$ MeV & 15-20 & 215.3 & 7.11$\pm$0.10 &
7.17$\pm$0.07 & 0.70 & 2275.5 & 10.1 & 3203.0 & 0.48
\\
\protect\cite{Rafelski:2004dp} & 20-30 & 166.6 & 6.45$\pm$0.09 &
6.47$\pm$0.07 & 0.71 & 1689.5 & 9.1 & 2386.3 & 0.58
\\
 & 30-40 & 114.2 & 5.57$\pm$0.08 & 5.63$\pm$0.06 & 0.70 & 1097.2 & 7.9 & 1544.0 & 0.77
\\
 & 40-50 & 74.4 & 4.68$\pm$0.07 & 4.85$\pm$0.06 & 0.69 & 669.0 & 6.7 & 929.8 & 1.05
\\
 & 50-60 & 45.5 & 3.83$\pm$0.07 & 4.16$\pm$0.05 & 0.68 & 383.9 & 5.7 & 521.6 & 1.13
\\
 & 60-70 & 25.7 & 2.99$\pm$0.06 & 3.47$\pm$0.05 & 0.65 & 194.3 & 4.6 & 256.5 & 1.41
\\
 & 70-80 & 13.4 & 2.22$\pm$0.06 & 2.78$\pm$0.05 & 0.62 & 86.3 & 3.6 & 110.4 & 1.55
\\
 & 80-92 & 6.3 & 1.71$\pm$0.06 & 2.40$\pm$0.05 & 0.58 & 44.2 & 2.9 & 54.2 & 1.40
\\
\hline STAR at & 0-5 & 352 & 9.22$\pm$0.18 & 7.13$\pm$0.06 & 0.79
& 3805.9 & 11.7 & 6222.1 & 0.29
\\
$\sqrt{s_{NN}}=200$ GeV: & 5-10 & 299 & 8.40$\pm$0.17 &
6.83$\pm$0.06 & 0.78 & 3029.0 & 10.8 & 4800.7 & 0.27
\\
$T = 160.0$ MeV & 10-20 & 234 & 7.57$\pm$0.15 & 6.33$\pm$0.06 &
0.77 & 2277.0 & 9.9 & 3548.3 & 0.23
\\
$\mu_{B} = 24.0$ MeV & 20-30 & 166 & 6.50$\pm$0.14 & 5.86$\pm$0.06
& 0.74 & 1557.9 & 8.8 & 2326.6 & 0.30
\\
\protect\cite{Barannikova:2005rw} & 30-40 & 115 & 5.52$\pm$0.12 &
5.37$\pm$0.06 & 0.72 & 1028.9 & 7.7 & 1476.2 & 0.27
\\
 & 40-50 & 76 & 4.66$\pm$0.11 & 4.91$\pm$0.06 & 0.69 & 669.8 & 6.8 & 922.7 & 0.27
\\
 & 50-60 & 47 & 3.87$\pm$0.10 & 4.40$\pm$0.06 & 0.66 & 413.3 & 5.9 & 550.2 & 0.35
\\
 & 60-70 & 27 & 3.07$\pm$0.09 & 3.94$\pm$0.06 & 0.61 & 232.9 & 5.0 & 295.3 & 0.46
\\
 & 70-80 & 14 & 2.37$\pm$0.08 & 3.32$\pm$0.06 & 0.58 & 116.8 & 4.1 & 143.4 & 0.87
\\
\hline BRAHMS at & 0-5 & 357 & 8.75$\pm$0.16 & 8.38$\pm$0.11 &
0.72 & 4029.6 & 12.1 & 5825.3 & 0.50
\\
$\sqrt{s_{NN}}=200$ GeV: & 0-10 & 328 & 8.50$\pm$0.15 &
8.08$\pm$0.10 & 0.72 & 3670.6 & 11.7 & 5329.1 & 0.52
\\
$T = 155.2$ MeV & 10-20 & 239 & 7.52$\pm$0.13 & 7.28$\pm$0.09 &
0.72 & 2584.1 & 10.5 & 3714.1 & 0.46
\\
$\mu_{B} = 26.4$ MeV & 20-40 & 140 & 6.29$\pm$0.12 & 6.20$\pm$0.09
& 0.71 & 1541.5 & 8.8 & 2194.7 & 0.36
\\
\protect\cite{Rafelski:2004dp} & 40-60 & 62 & 4.42$\pm$0.12 &
4.95$\pm$0.10 & 0.67 & 608.2 & 6.6 & 816.0 & 0.61
\\
\hline
\end{tabular}
\end{ruledtabular}
\end{table}

The final results for the geometric parameters $\rho_{max}$ and
$\tau$ are gathered in Table~\ref{Table1} together with the
corresponding values of $\chi^{2}$/NDF for each centrality class
additionally characterized by the number of participants
$N_{part}$. Note that besides the most peripheral bins of the
PHENIX measurements all fits are statistically significant. Other
physical quantities like the surface velocity
$\beta_{\perp}^{max}$, the volume at the beginning of the
freeze-out $V_{f.o.}^{(1)}$, the maximal freeze-out time at the
central slice $t_{f.o.}^{(2)}$ and the corresponding volume
$V_{f.o.}^{(2)}$, which have been defined in Sec.~\ref{Foundat},
are also given there. Note that values of $\rho_{max}$ and $\tau$
fitted for the same spectra as in \cite{Broniowski:2002nf} (the
$0-5 \%$, $15-30 \%$ and $60-92 \%$ centrality bins for PHENIX at
$\sqrt{s_{NN}}=130$ GeV) are the same with $5 \%$ accuracy as
given in \cite{Broniowski:2002nf}, besides the peripheral bin. The
model calculations take into account full feeding from weak decays
besides the PHENIX case at $\sqrt{s_{NN}}=200$ GeV where protons
(antiprotons) from $\Lambda$ ($\bar{\Lambda}$) decays are
excluded. This needs some comments because the STAR Collaboration
claims that its pion spectra are corrected for weak decays
\cite{Adams:2003xp}. However, in the present analysis the best
quality of the fit expressed by the value of $\chi^{2}$/NDF in the
STAR case is for full feeding from weak decays. Namely, if pions
from $\Lambda$ ($\bar{\Lambda}$) decays are subtracted then
$\chi^{2}$/NDF = 0.37 for the $0-5 \%$ centrality bin, whereas if
pions from all possible weak decays are excluded $\chi^{2}$/NDF =
0.88 for the same bin. With the full inclusion of pions from weak
decays $\chi^{2}$/NDF = 0.29 (see Table~\ref{Table1}). That the
single-freeze-out model with all the resonances and the full
feeding from weak decays can fit the STAR spectra has been already
noticed in \cite{Adams:2003xp} but, as it is stated there, with
not satisfactory $\chi^{2}$/NDF.

\begin{figure}
\includegraphics[width=0.45\textwidth]{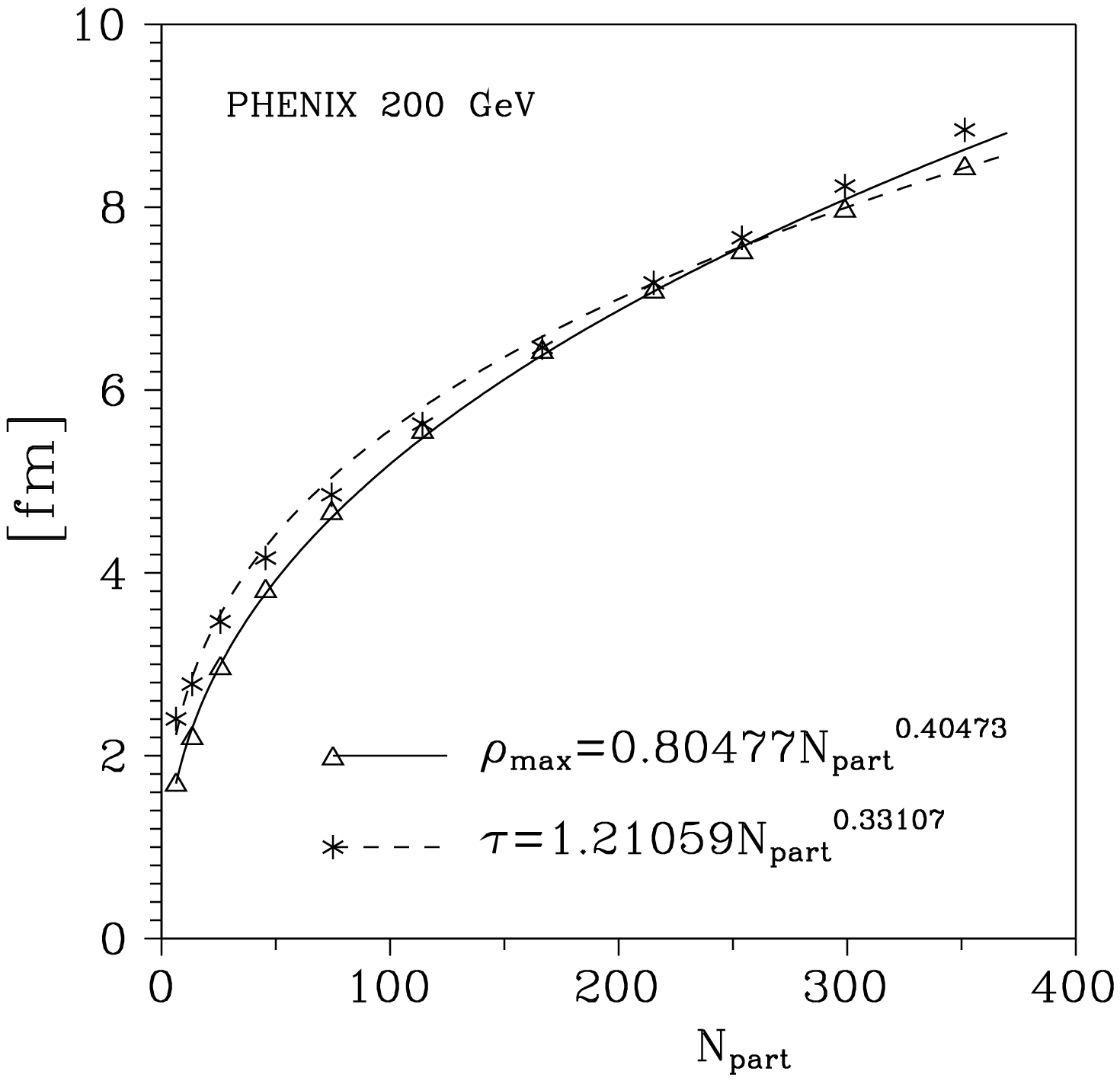}
\caption{\label{Fig.1} Values of the geometric parameters of the
model from the fourth and fifth column of Table~\ref{Table1} for
PHENIX at $\sqrt{s_{NN}}=200$ GeV. The lines are the best power
approximations. }
\end{figure}

\begin{figure}
\includegraphics[width=0.45\textwidth]{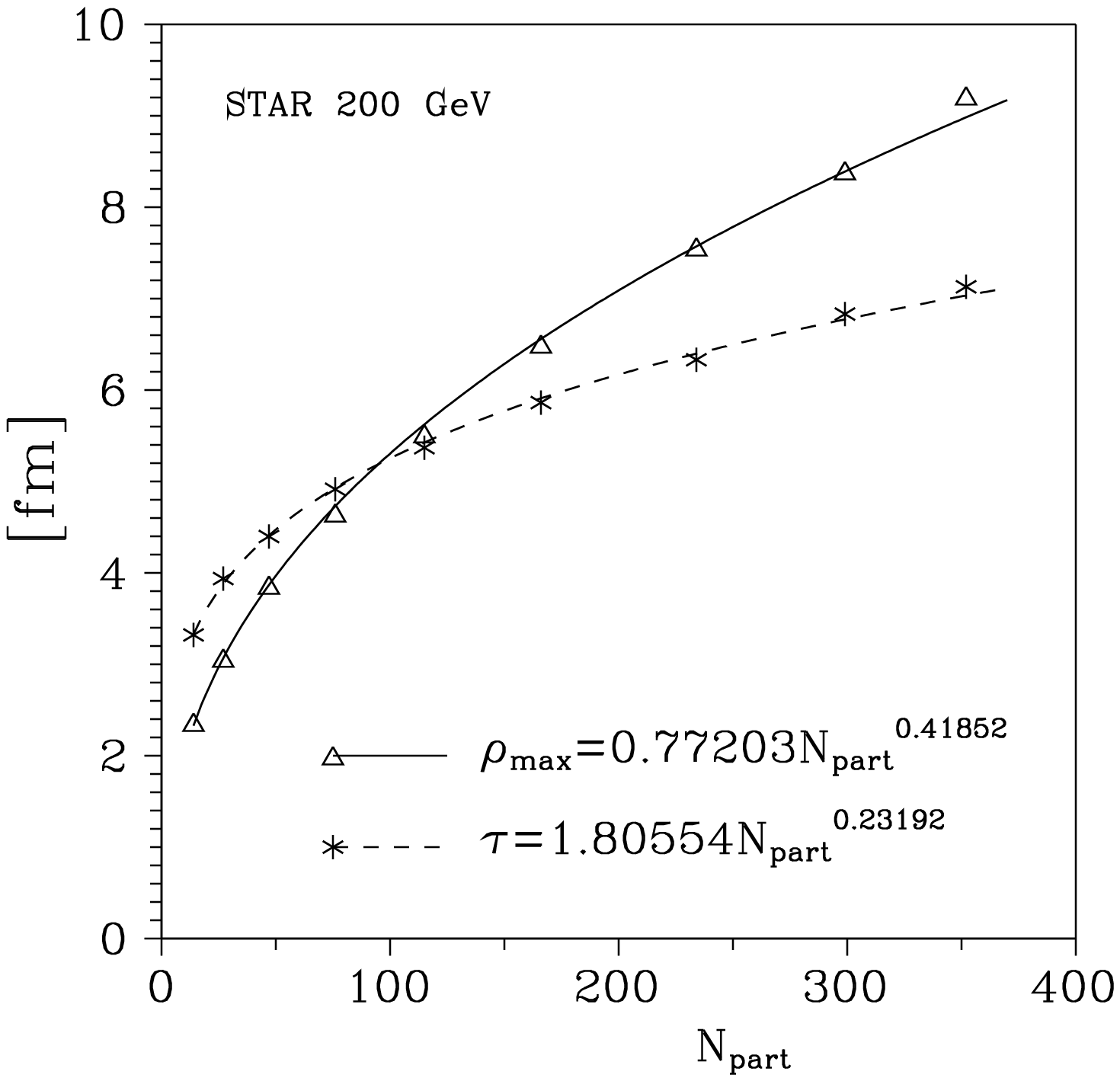}
\caption{\label{Fig.2} Values of the geometric parameters of the
model from the fourth and fifth column of Table~\ref{Table1} for
STAR at $\sqrt{s_{NN}}=200$ GeV. The lines are the best power
approximations. }
\end{figure}

\begin{figure}
\includegraphics[width=0.45\textwidth]{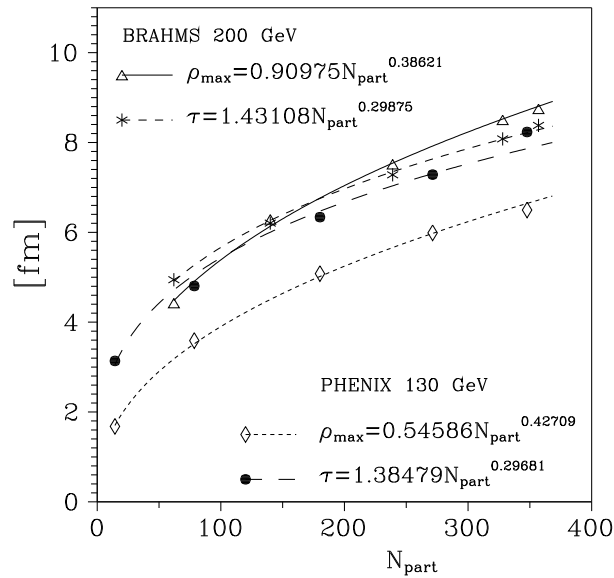}
\caption{\label{Fig.3} Values of the geometric parameters of the
model from the fourth and fifth column of Table~\ref{Table1} for
BRAHMS at $\sqrt{s_{NN}}=200$ GeV and PHENIX at
$\sqrt{s_{NN}}=130$ GeV. The lines are the best power
approximations. }
\end{figure}

Values of the geometric parameters $\rho_{max}$ and $\tau$ from
Table~\ref{Table1} are presented in Figs.\,\ref{Fig.1}-\ref{Fig.3}
as functions of $N_{part}$. Also there the lines of the best power
approximations are depicted,

\begin{equation}
x \sim  N_{part}^{\kappa},\;\;\;\;\;\;\;\; x=\rho_{max},\;\tau,
\label{Scalbeha}
\end{equation}

\noindent with a scaling exponent $\kappa = 0.40-0.43$ for
$\rho_{max}$ and $\kappa = 0.23-0.33$ for $\tau$. Note that the
scaling is almost ideal, only slight deviations can be observed
for the most central bins in some cases ($\rho_{max}$ and $\tau$
for PHENIX at $\sqrt{s_{NN}}=200$ GeV, Fig.\,\ref{Fig.1}, and
$\tau$ for PHENIX at $\sqrt{s_{NN}}=130$ GeV, Fig.\,\ref{Fig.3}).
Another interesting point would be to give at least a rough notion
of what the predicted volumes $V_{f.o.}^{(1)}$ and
$V_{f.o.}^{(2)}$ mean. The simplest way is to compare them with
the volume of the colliding nucleus, which is the gold nucleus
here. Since the volume of the gold nucleus at rest ($V_{Au}$)
equals about 1160 fm$^{3}$ ($V_{A} = 4/3\;\pi r_{0}^{3}A,\;
r_{0}=1.12$ fm), one can see that for the most central collisions
at $\sqrt{s_{NN}}=200$ GeV the volume at the beginning of the
freeze-out, $V_{f.o.}^{(1)}$, is $\approx 3V_{Au}$, whereas the
volume when the freeze-out ceases at the central slice,
$V_{f.o.}^{(2)}$, equals $\approx 5V_{Au}$. For the most central
bin at $\sqrt{s_{NN}}=130$ GeV this is $\approx 1.9V_{Au}$ and
$\approx 2.4V_{Au}$, respectively. As far as the lifespan of the
system is considered the final moment of the freeze-out in the
central slice, $t_{f.o.}^{(2)}$, has turned out to be about 10 fm
for the most central bins. With the realistic assumption of the
maximal possible value of the fluid rapidity,
$\alpha_{\parallel}^{max}$, which will be explored in
Sec.~\ref{Totalcharge}, and for values of
$\alpha_{\parallel}^{max}$ obtained there, the moment when the
freeze-out ceases entirely, $t_{f.o.}^{(3)}$, will be of the order
of 10 $t_{f.o.}^{(2)}$, that is of the order of 100 fm for the
most central collisions.


\begin{figure}
\includegraphics[width=0.45\textwidth]{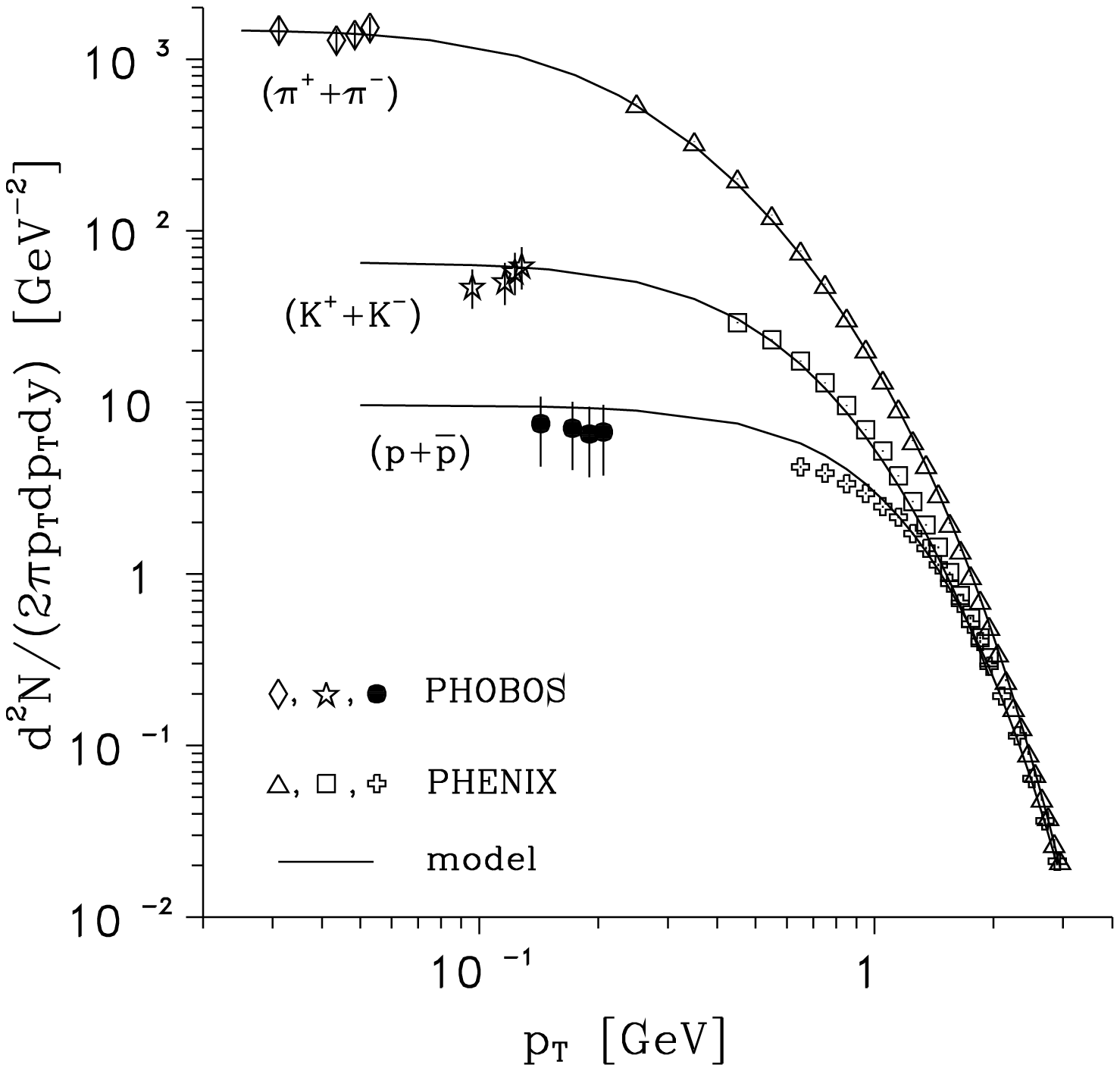}
\caption{\label{Fig.4} Invariant yields as a function of $p_{T}$
for RHIC at $\sqrt{s_{NN}}=200$ GeV. The PHOBOS data are for the
$15 \%$ most central collisions with the error bars expressed as
the sum of the systematic and statistical uncertainties
\protect\cite{Back:2004zx}. The corresponding PHENIX data are
presented as the averages of the invariant yields for the $0-5
\%$, $5-10 \%$ and $10-15 \%$ centrality bins with no errors
given. Lines are the appropriate predictions of the
single-freeze-out model. }
\end{figure}

The extension of the model predictions for the invariant
distributions to the low-$p_{T}$ region (0.03-0.05 GeV for pions,
0.09-0.13 GeV for kaons and 0.14-0.21 GeV for protons and
antiprotons) agree well with the available PHOBOS data
\cite{Back:2004zx}, even protons and antiprotons agree within
errors as one can see in Fig.\,\ref{Fig.4} (but the $\approx 30
\%$ overestimation of protons and antiprotons has been obtained
for the lowest $p_{T}$ point of the PHENIX data, \textit{i.e.} for
$p_{T}=0.65$ GeV). The similar result was obtained also for the
preliminary data but protons and antiprotons were much more
overweight \cite{Baran:2003nm}. In Fig.\,\ref{Fig.4} the original
PHENIX data \cite{Adler:2003cb} have been modified so as to be
more useful for the comparison with the PHOBOS data
\cite{Back:2004zx}. Namely, the PHOBOS measurements were done for
the $15 \%$ most central collisions whereas the PHENIX ones for
the $0-5 \%$, $5-10 \%$ and $10-15 \%$ centrality bins. Since the
treatment of counts includes the averaging over the number of
events in a given centrality bin and for the same run the number
of events in the $15 \%$ most central bin should be equal to the
sum of numbers of events in the $0-5 \%$, $5-10 \%$ and $10-15 \%$
centrality bins, the rough approximation of the hypothetic
measurement done in the $0-15 \%$ centrality bin would be the
average of the measurements done in the $0-5 \%$, $5-10 \%$ and
$10-15 \%$ centrality bins. Therefore, in Fig.\,\ref{Fig.4} the
data points representing the PHENIX yields are such averages of
the original values from \cite{Adler:2003cb}. The model
predictions have been averaged in the same way.

\subsection {Estimations of transverse energy and charged particle multiplicity
densities} \label{Finletnch}

\begin{figure}
\includegraphics[width=0.45\textwidth]{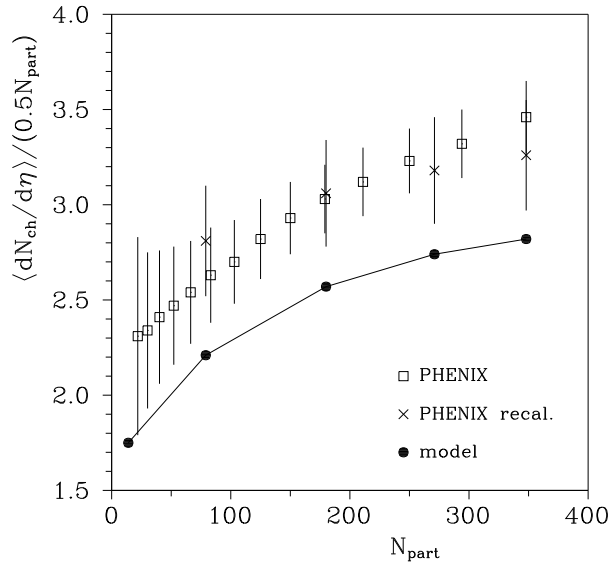}
\caption{\label{Fig.5} $dN_{ch}/d\eta$ per pair of participants
versus $N_{part}$ for RHIC at $\sqrt{s_{NN}}=130$ GeV. The PHENIX
original data are from \protect\cite{Adler:2004zn}. Also the
recalculated PHENIX data from summing up the integrated charged
hadron yields delivered in \protect\cite{Adcox:2003nr} are
depicted. The line connects the results and is to guide the eye.}
\end{figure}

\begin{figure}
\includegraphics[width=0.45\textwidth]{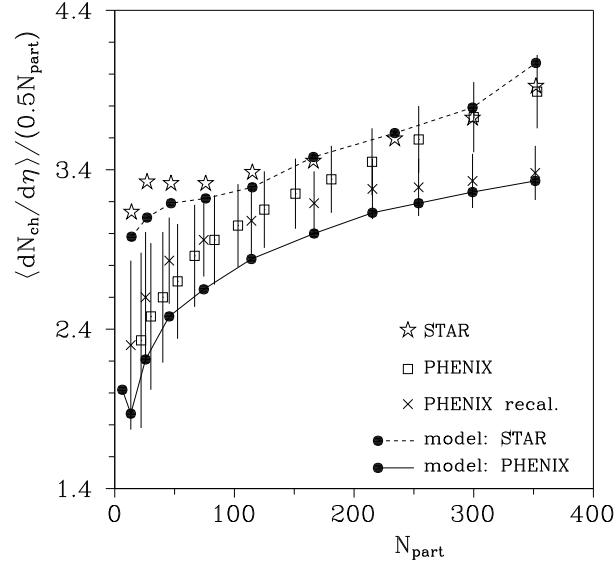}
\caption{\label{Fig.6} $dN_{ch}/d\eta$ per pair of participants
versus $N_{part}$ for RHIC at $\sqrt{s_{NN}}=200$ GeV. The
original PHENIX data are from \protect\cite{Adler:2004zn}, whereas
the recalculated PHENIX data are from summing up the integrated
charged hadron yields delivered in \protect\cite{Adler:2003cb}.
Also the STAR data are depicted with no errors given as in the
source paper \protect\cite{Adams:2003xp}. The lines connect the
results and are to guide the eye. }
\end{figure}

The results of numerical estimations of $dN_{ch}/d\eta\vert_{mid}$
divided by the number of participant pairs for various centrality
classes are presented in Figs.\,\ref{Fig.5} and \ref{Fig.6} for
RHIC at $\sqrt{s_{NN}}=130$ and 200 GeV, respectively.
Additionally to the straightforward PHENIX measurements of the
charged particle multiplicity density, also the data from the
summing up of the integrated charged hadron yields
\cite{Adler:2003cb,Adcox:2003nr} are depicted in these figures
(more precisely, since the integrated charged hadron yields are
given as rapidity densities, the transformation to pseudo-rapidity
should be done, which means the division by a factor 1.2 here, see
\cite{Bazilevsky:2002fz}). Also in \cite{Adler:2003cb} the feeding
of $p(\bar{p})$ from $\Lambda(\bar{\Lambda})$ decays is excluded.
To diminish this effect, integrated $p$ and $\bar{p}$ yields
delivered in \cite{Adler:2003cb} were corrected to include back
the feeding. The correction was done by the division by a factor
0.65, which is the rough average of a $p_{T}$-dependent multiplier
used by PHENIX Collaboration (see Fig.8 and Eq.(5) in
\cite{Adler:2003cb}). Generally, the agreement of the model
predictions with the data is much better for RHIC at
$\sqrt{s_{NN}}=200$ GeV. For the case of $\sqrt{s_{NN}}=130$ GeV,
only the qualitative agreement has been reached. The main reason
of the $\approx 15 \%$ underweight of $dN_{ch}/d\eta$ seen in
Fig.\,\ref{Fig.5} is that the best fit estimates for the $p_{T}$
spectra of pions at the lowest $p_{T}$ lies $29 \%$ on the average
below the corresponding measured values. This concerns also the
PHENIX case at $\sqrt{s_{NN}}=200$ GeV but the underestimation of
low-$p_{T}$ pions is smaller there ($15 \%$ on the average). Since
pions comprise about $80 \%$ of $dN_{ch}/d\eta$ and their
low-$p_{T}$ fraction can contribute even 30 \% to the integrated
yield \cite{Adcox:2001mf}, this explains the obtained underweight
of $dN_{ch}/d\eta$ in spite of the very good quality of the fit
for PHENIX at $\sqrt{s_{NN}}=130$ GeV. The discrepancy between the
directly measured $dN_{ch}/d\eta$ and $dN_{ch}/d\eta$ expressed as
the sum of the integrated charged hadron yields can be the next
reason, especially for RHIC at $\sqrt{s_{NN}}=200$ GeV (see
Fig.\,\ref{Fig.6}; this effect was notified in backup slides of
\cite{Chujo:2002bi}). The discrepancy starts at mid-centrality and
rises with the centrality. The another reason for the quantitative
disagreement is that transverse momentum spectra are measured in
\emph{limited ranges}, so very important low-$p_{T}$ regions are
not covered by the data. To obtain integrated yields some
extrapolations below and above the measured ranges are used. In
fact these extrapolations are only analytical fits, but
contributions from regions covered by them account for about
$25-40\%$ of the integrated yields \cite{Adcox:2001mf}. It might
turned out that these extrapolations differ from the thermal
distributions supplemented by the distributions of products of
decays. However, the extension of the model predictions for the
invariant distributions to the low-$p_{T}$ region agree well with
the available data, \textit{cf.} Fig.\,\ref{Fig.4}.

\begin{figure}
\includegraphics[width=0.45\textwidth]{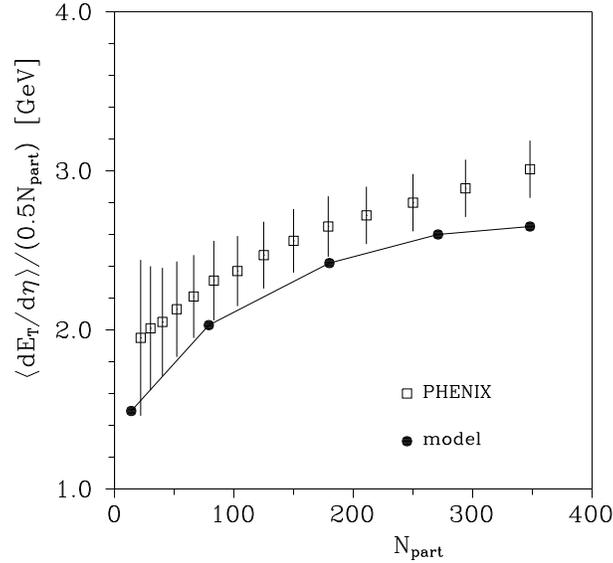}
\caption{\label{Fig.7} $dE_{T}/d\eta$ per pair of participants
versus $N_{part}$ for RHIC at $\sqrt{s_{NN}}=130$ GeV. The PHENIX
data are from \protect\cite{Adler:2004zn}. The line connects the
results and is to guide the eye. }
\end{figure}

\begin{figure}
\includegraphics[width=0.45\textwidth]{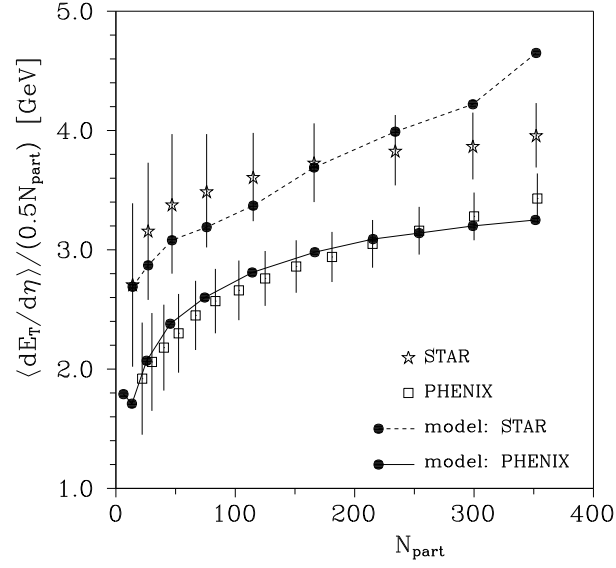}
\caption{\label{Fig.8} $dE_{T}/d\eta$ per pair of participants
versus $N_{part}$ for RHIC at $\sqrt{s_{NN}}=200$ GeV. The PHENIX
data are from \protect\cite{Adler:2004zn} but the original STAR
data from \protect\cite{Adams:2004cb} have been rescaled to
$\eta=0$, see text for more explanations). The lines connect the
results and are to guide the eye. }
\end{figure}

The values of $dE_{T}/d\eta\vert_{mid}$ divided by the number of
participant pairs are shown in Figs.\,\ref{Fig.7} and \ref{Fig.8}
for $\sqrt{s_{NN}}=130$ and 200 GeV, respectively. The quality of
the model predictions for $dE_{T}/d\eta$ is much better then for
$dN_{ch}/d\eta$. Again the worse agreement is for the case of
$\sqrt{s_{NN}}=130$ GeV. The STAR measurements need separate
comments. The STAR data were taken not at mid-rapidity ($\eta=0$)
as in the PHENIX case but at $\eta=0.5$ \cite{Adams:2004cb}. But
the $p_{T}$ spectra used for fits of the geometric parameters were
measured at mid-rapidity, also in the STAR case
\cite{Adams:2003xp}. Therefore one should expect some
overestimation of the predictions in comparison with the data on
the whole. To remove this effect the original STAR data
\cite{Adams:2004cb} have been divided by a factor
$\sin{(\theta\vert_{\eta=0.5})}\approx 0.887$. As it can be seen
from Fig.\,\ref{Fig.8}, the predictions and data agree with each
other within errors besides two most central points. But even
there, the discrepancy is $\approx 17\%$. It should be also
noticed that bigger values of transverse energy estimates for the
STAR case are the consequence of slightly higher values of $p_{T}$
distributions measured by STAR Collaboration with respect to
PHENIX measurements (it can be seen directly from careful
comparison of spectra given in \cite{Adams:2003xp} and
\cite{Adler:2003cb}).


\begin{figure}
\includegraphics[width=0.45\textwidth]{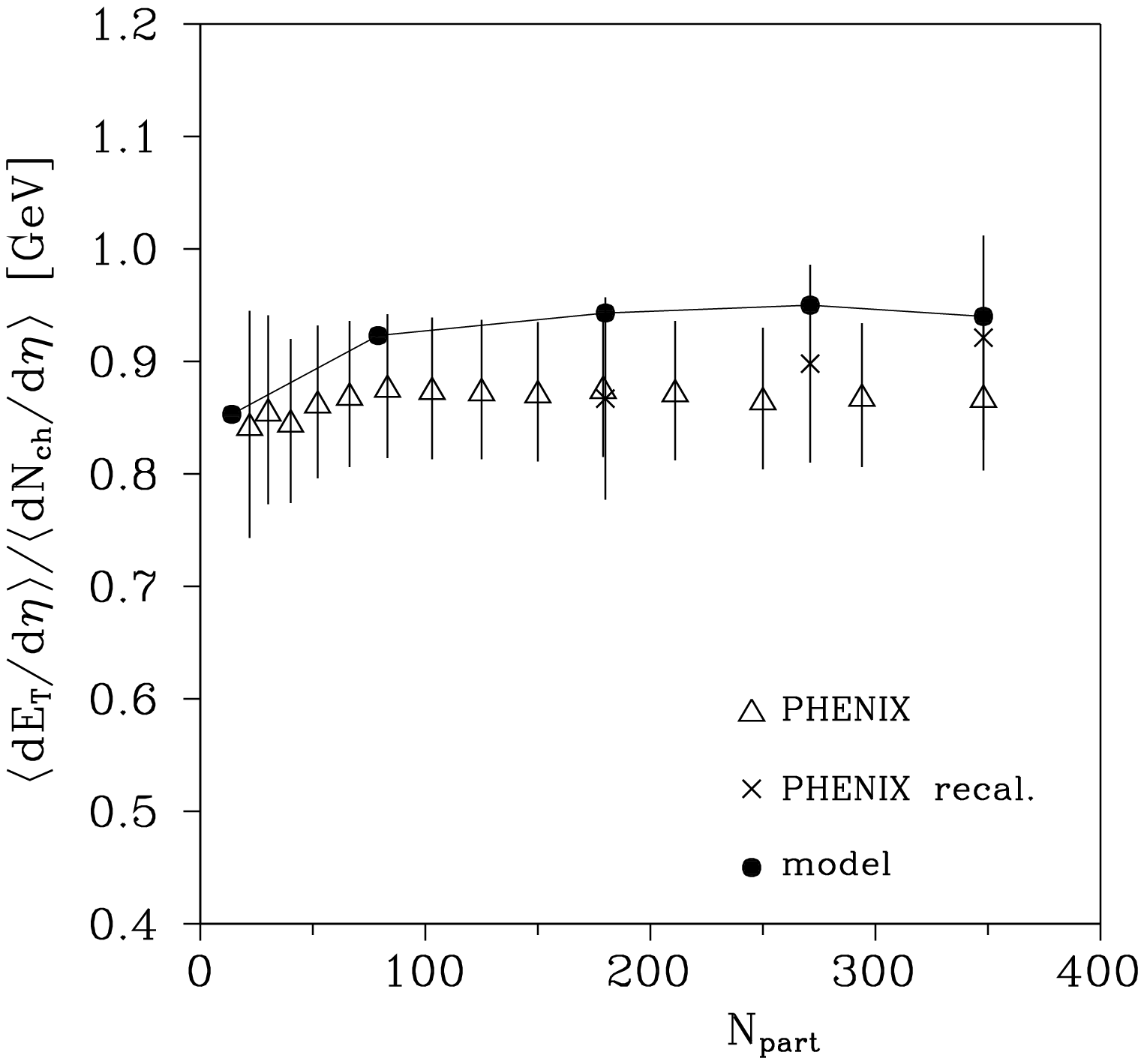}
\caption{\label{Fig.9} $\langle dE_{T}/d\eta\rangle /\langle
dN_{ch}/d\eta\rangle$ versus $N_{part}$ for RHIC at
$\sqrt{s_{NN}}=130$ GeV. The original PHENIX data are from
\protect\cite{Adler:2004zn}. The recalculated PHENIX data are also
depicted, here "recalculated" means that the sum of integrated
charged hadron yields \cite{Adcox:2001mf} have been substituted
for the denominator in the ratio. The line connects the results
and is to guide the eye. }
\end{figure}
\begin{figure}
\includegraphics[width=0.45\textwidth]{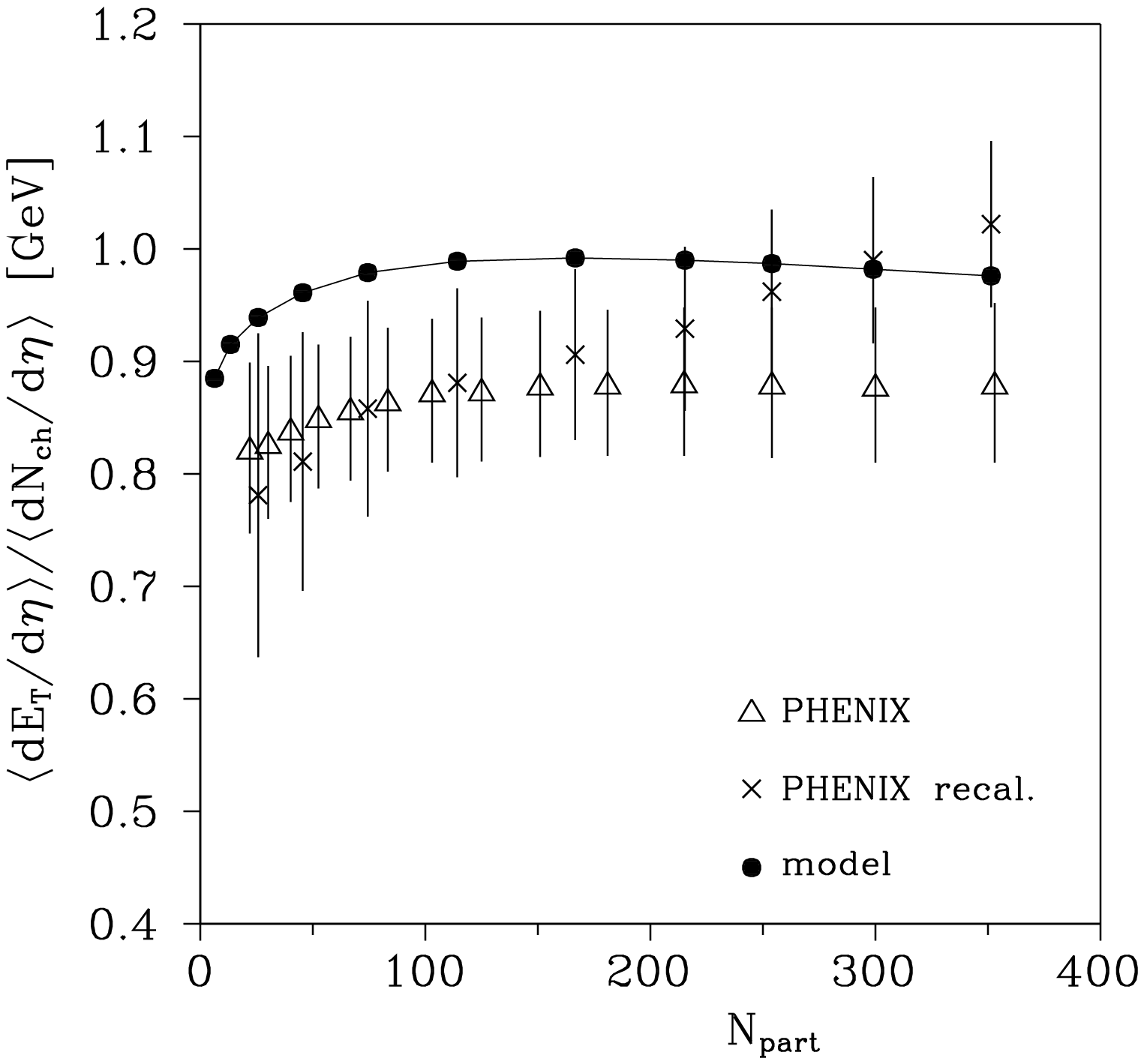}
\caption{\label{Fig.10} $\langle dE_{T}/d\eta\rangle /\langle
dN_{ch}/d\eta\rangle$ versus $N_{part}$ for RHIC at
$\sqrt{s_{NN}}=200$ GeV. The original PHENIX data are from
\protect\cite{Adler:2004zn}. The recalculated PHENIX data are also
depicted, here "recalculated" means that the sum of integrated
charged hadron yields \cite{Adler:2003cb} have been substituted
for the denominator in the ratio. The line connects the results
and is to guide the eye. }
\end{figure}
\begin{figure}
\includegraphics[width=0.45\textwidth]{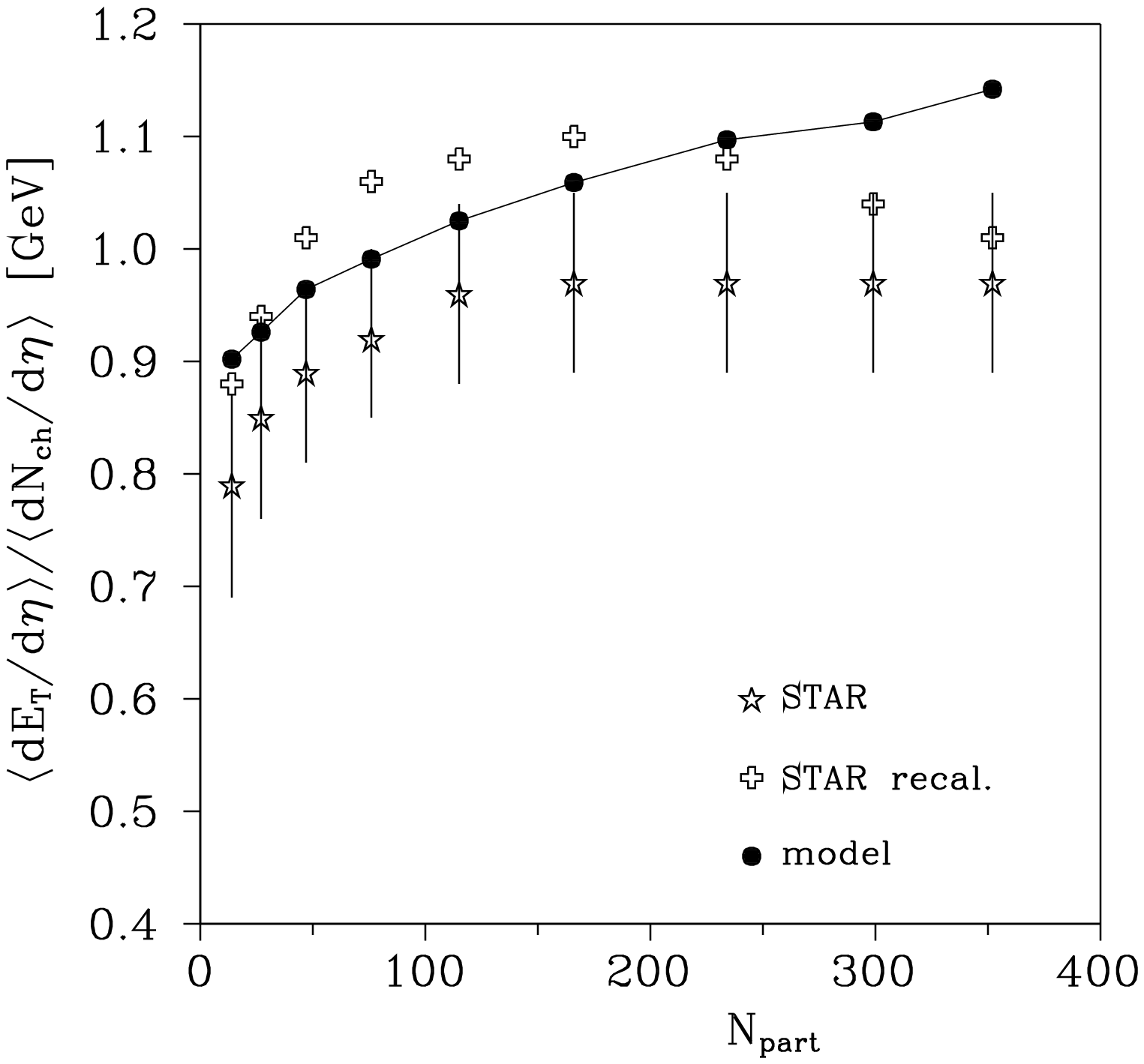}
\caption{\label{Fig.11} $\langle dE_{T}/d\eta\rangle /\langle
dN_{ch}/d\eta\rangle$ versus $N_{part}$ for RHIC at
$\sqrt{s_{NN}}=200$ GeV. The original STAR data
\protect\cite{Adams:2004cb} have been rescaled to $\eta=0$ by the
division by a factor $0.887$. Also the recalculated STAR data are
depicted, here "recalculated" means that the original values of
$dE_{T}/d\eta$ \protect\cite{Adams:2004cb} have been divided by a
factor $0.887$ and values of $dN_{ch}/d\eta$ measured at
midrapidity \protect\cite{Adams:2003xp} are taken for the
denominator, however errors are not given in
\protect\cite{Adams:2003xp}. The line connects the results and is
to guide the eye. }
\end{figure}

Values of the ratio $\langle dE_{T}/d\eta\rangle /\langle
dN_{ch}/d\eta\rangle$ as a function of $N_{part}$ are presented in
Figs.\,\ref{Fig.9}-\ref{Fig.11}. As one can see, the position of
model predictions is very regular and exactly resembles the
configuration of the data in each case, the estimates are only
shifted up about $\approx 10 \%$ as a whole. For the PHENIX case,
Figs.\,\ref{Fig.9}-\ref{Fig.10}, the shift is about $8 \%$ at
$\sqrt{s_{NN}}=130$ GeV and about $14 \%$ at $\sqrt{s_{NN}}=200$
GeV. This overestimation of the ratio can be explained, at least
for more central collisions, by the observed discrepancy between
the directly measured $dN_{ch}/d\eta$ and $dN_{ch}/d\eta$
expressed as the sum of the integrated charged hadron yields. If
the original data points are replaced by the recalculated data,
such that the denominators are sums of the integrated charged
hadron yields, then much better agreement can be reached for more
central collisions. In the STAR case, Fig.~\ref{Fig.11}, as the
experimental data at $\eta=0$ two modified versions of the
original data \cite{Adams:2004cb} at $\eta=0.5$ are depicted. The
first version was done by the division of the original data by a
factor $\sin{(\theta\vert_{\eta=0.5})}\approx0.887$. In the second
one, besides the above-mentioned rescaling, the values of
$dN_{ch}/d\eta$ measured directly at $\eta=0$ \cite{Adams:2003xp}
were put into the denominator. Up to the mid-centrality the model
estimates agree pretty well with the second version of the data,
but for the most central bin the overestimation of the ratio of
about $13 \%$ has been obtained.

\subsection {Total charged-particle multiplicity}
\label{Totalcharge}

Generally, for any expansion satisfying the condition
$d\sigma_{\mu} \sim u_{\mu}$ on a freeze-out hypersurface, the
total multiplicity of particle species $i$ can be derived in the
form (for the more formal proof see \cite{Broniowski:2002nf})

\begin{eqnarray}
N_{i} &=& \int d^{2}p_{T}\;dy\;{{dN_{i}} \over {d^{2}p_{T}\;dy}}=
\int d^{2}p_{T}\;dy \int p^{\mu}d\sigma_{\mu}\;f_{i}(p \cdot u) =
\int d\sigma \int d^{2}p_{T}\;dy\;(p \cdot u)\;f_{i}(p \cdot u)
\cr \cr && = \int d\sigma \int { d^{3}\vec{p} \over E}\;(p \cdot
u)\;f_{i}(p \cdot u)= \int d\sigma \; n_{i}(T,\mu_{B}) =
n_{i}(T,\mu_{B})\int d\sigma \;, \label{Totmult}
\end{eqnarray}

\noindent if the local thermal parameters are constant on this
hypersurface. This means that under such condition the total
multiplicity is a product of the density $n_{i}$ calculated for a
static gas case and the hypersurface volume. Note that $n_{i}$ is
not the primordial thermal density of particle species $i$ but it
collects also contributions from decays of resonances. Since the
freeze-out hypersurface and the flow described in
Sec.~\ref{Foundat} fulfill the condition $d\sigma_{\mu} \sim
u_{\mu}$, the formula (\ref{Totmult}) is valid in this case as
well. In practise the rapidity of the fluid element
$\alpha_{\parallel}$ should not be unlimited but should have its
maximal value $\alpha_{\parallel}^{max}$. Otherwise, as it can be
seen further, the hypersurface volume and the total charged
particle multiplicity would be infinite. Then, with the use of
Eq.~(\ref{Cooper3}) one can express the hypersurface volume as

\begin{equation}
\int d\sigma = \tau^{3}\;
\int\limits_{-\alpha_{\parallel}^{max}}^{+\alpha_{\parallel}^{max}}
d\alpha_{\parallel}\;\int\limits_{0}^{\rho_{max}/\tau}\;\sinh{\alpha_{\perp}}\;
d(\sinh{\alpha_{\perp}})\; \int\limits_{0}^{2\pi} d\xi =
2\pi\;\alpha_{\parallel}^{max} \tau \rho_{max}^{2}
\;.\label{Hypvolum}
\end{equation}

\noindent Finally, the total multiplicity of charged particles can
be obtained:

\begin{equation}
N_{ch} = 2\pi\;\alpha_{\parallel}^{max} \tau \rho_{max}^{2}
\sum_{i \in B} n_{i}(T,\mu_{B}) = 2\pi\;\alpha_{\parallel}^{max}
\tau \rho_{max}^{2}\; n_{ch}(T,\mu_{B}) \;.\label{Totcharged}
\end{equation}

\noindent One can see that the above formula has the form of the
product of a volume and a density. This volume can be treated as
an approximation of the size of the gas at the freeze-out (or an
average volume of a hadron source). Approximation - because it is
the volume of a static fireball which would emit the same amount
of charged hadrons as the freeze-out hypersurface described by
Eq.~(\ref{Hypsur}) with the superimposed limit in the range of
$\alpha_{\parallel}$. Average - because for the static fireball
the freeze-out happens at a given moment of time, whereas in the
discussed model the freeze-out extends in time. This approximate
volume will be called $V_{stat.}$ from now on and equals

\begin{equation}
V_{stat.}=2\pi\;\alpha_{\parallel}^{max} \tau \rho_{max}^{2} \;.
\label{Volstatap}
\end{equation}

For $\alpha_{\parallel}^{max}$ the following reasonable assumption
has been made: it is equal to the rapidity of leading baryons
after the collision. In other words, it is assumed that the fluid
which has been created in the CRR can not move faster in the
longitudinal direction then fragments of a target or a projectile
after the collision. Therefore it should depend on the centrality
of the collision, since the more central the collision is, the
higher degree of the stopping of the initial baryons ought to
happen in principle. There are two limiting cases, the maximum
stopping happens for the most central collision whereas if the
centrality approaches $100 \%$ the stopping disappears. For values
of $\alpha_{\parallel}^{max}$ it means that for the most central
collision $\alpha_{\parallel}^{max} = \langle y \rangle = y_{p} -
\langle \delta y \rangle$ and in the limiting case of the $100 \%$
centrality $\alpha_{\parallel}^{max} =  y_{p}$, where $\langle y
\rangle$ is the mean net-baryon rapidity after the collision,
$y_{p}$ the projectile rapidity and $\langle \delta y \rangle$ the
average rapidity loss \cite{Bearden:2003hx}. And the last
assumption is that $\alpha_{\parallel}^{max}$ is a linear function
of the centrality $c$, where $c$ is a fractional number
representing the middle of a given centrality bin, \emph{i.e.}
$c=0.025$ for the $0-5 \%$ centrality bin, $c=0.075$ for the $5-10
\%$ centrality bin, etc.. Then $\alpha_{\parallel}^{max}(c)$ has
the following form:

\begin{equation}
\alpha_{\parallel}^{max}(c) = y_{p} - { \langle \delta y \rangle
\over 0.975 } \cdot (1-c) \;.\label{Alfpmax}
\end{equation}

\begin{figure}
\includegraphics[width=0.45\textwidth]{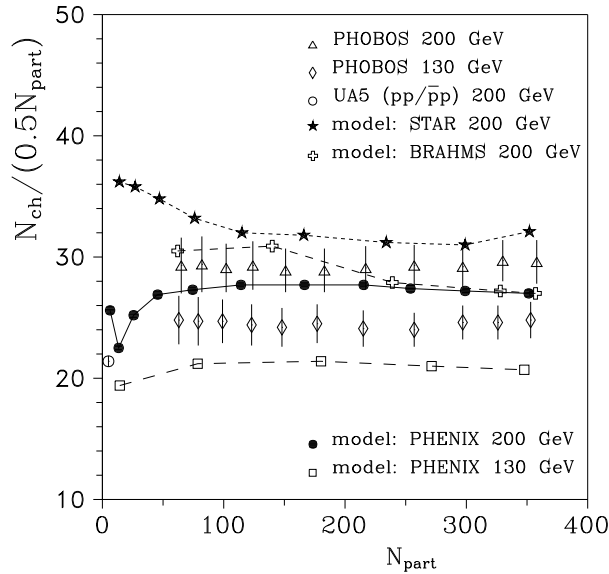}
\caption{\label{Fig.12} $N_{ch}$ per pair of participants versus
$N_{part}$ for RHIC at $\sqrt{s_{NN}}=130$ and $200$ GeV. The
PHOBOS data are from \protect\cite{Back:2003xk} and the
$pp/\bar{p}p$ data point of the UA5 measurement is from Fig.39.5
in \protect\cite{Hagiwara:fs}. The lines connect the results and
are to guide the eye. }
\end{figure}

\noindent In the derivation of Eq.~(\ref{Alfpmax}) the most
central bin has been taken as the $0-5 \%$ centrality bin, since
the estimate of the average rapidity loss $\langle \delta y
\rangle$ for RHIC at $\sqrt{s_{NN}}= 200$ GeV was done for this
centrality class in \cite{Bearden:2003hx}. The BRAHMS
Collaboration reports $\langle \delta y \rangle = 2.05$ for RHIC
at $\sqrt{s_{NN}}= 200$ GeV ($y_{p} = 5.36$)
\cite{Bearden:2003hx}. Since for RHIC at $\sqrt{s_{NN}}= 130$ GeV
the average rapidity loss has not been given yet, it is assumed
that $\langle \delta y \rangle$ behaves linearly in $y_{p}$
between the SPS point (the $5 \%$ most central Pb-Pb collision at
158 GeV per nucleon: $y_{p} = 2.9$ in the c.m.s. and $\langle
\delta y \rangle = 1.76$ \cite{Appelshauser:1998yb}) and the
BRAHMS one. This results in $\langle \delta y \rangle = 2.0$ for
RHIC at $\sqrt{s_{NN}}= 130$ GeV ($y_{p} = 4.93$). The numerical
estimates of $n_{ch}(T,\mu_{B})$ for the static gas give
$n_{ch}(T=155.2\;MeV,\;\mu_{B}=26.4\;MeV)= 0.36112$ fm$^{-3}$ and
$n_{ch}(T=160\;MeV,\;\mu_{B}=24\;MeV)= 0.44839$ fm$^{-3}$ for RHIC
at $\sqrt{s_{NN}}= 200$ GeV and $n_{ch}(T=165\;MeV,\;\mu_{B}=41\;
MeV)= 0.5618$ fm$^{-3}$ for RHIC at $\sqrt{s_{NN}}= 130$ GeV. Now
the total multiplicity of charged particles $N_{ch}$ can be
calculated with the use of Eqs.~(\ref{Totcharged}) and
(\ref{Alfpmax}). The results presented as the total
charged-particle multiplicity per participating pair versus
$N_{part}$ are gathered in Fig.\,\ref{Fig.12}. The predictions for
STAR and PHENIX exhibit almost ideal centrality independence of
the total charged-particle multiplicity per participating pair
within the range of the PHOBOS measurement, \textit{i.e.}
$N_{part} \approx 60-360$. For the BRAHMS case there is some
deviation from this ideal behavior, but it is of the order of $10
\%$. Additionally, the predicted values agree with the data within
$\approx 10 \%$.

\subsection {Comparison with the statistical hadronization model}
\label{Shmodel}

Another very interesting point would be the comparison of the
predictions of this model and the statistical hadronization model
(SHM) of Ref.~\cite{Rafelski:2004dp}, from where values of the
statistical parameters have been taken as an input to fit PHENIX
and BRAHMS spectra at $\sqrt{s_{NN}}=200$ GeV in
Sec.~\ref{Finlrhotau}. An indirect comparison has been already
done since good quality fits to the spectra have been obtained
with the substitution of SHM statistical parameter estimates. This
means that fits to the spectra with this model confirm values of
statistical parameters obtained from fits to particle yields with
SHM. The only left quantity which is independently obtained in
these both models and could be directly compared is the volume of
the hadron source. However not the global volume but the volume
fraction $dV/dy$ is predicted in \cite{Rafelski:2004dp}. The
volume fraction is defined by the relation
\cite{Letessier:2005qe}:

\begin{equation}
{ {dN_{i}} \over {dy} } = { {dV} \over {dy} } \cdot n_{i}\;,
\label{Volrapdef}
\end{equation}

\noindent which in principle should hold for each measured
particle species $i$. Note that $dN_{i}/dy$ depends on the
centrality, so $dV/dy$ does as well. According to Eq.~(5) in
\cite{Letessier:2005qe} the volume of the hadron source is given
by

\begin{equation}
V_{SHM} = k \; { {dV} \over {dy} } \cdot 2y_{p}\;,
\label{Volhaddef}
\end{equation}

\noindent where $k$ is a constant, $k \simeq 0.4-0.6$ for RHIC and
$2y_{p}$ expresses the maximum rapidity range. Since this volume
is in fact the volume of a static fireball, the most natural
volume appearing in the single freeze-out model which $V_{SHM}$
can be compared with is $V_{stat.}$ - the static equivalent of the
freeze-out volume, defined by Eq.~(\ref{Volstatap}). The detailed
comparison of these both estimates of the volume of the hadron
source for different centrality bins of PHENIX measurements at
$\sqrt{s_{NN}}=200$ GeV is given in Table~\ref{Table2}. One can
see that values of $V_{stat.}$ and $V_{SHM}$ agree with each other
qualitatively. Some overestimation of $V_{stat.}$ with respect to
$V_{SHM}$ can be observed and this effect increases with the
centrality. For the most central bin it is $\approx 6\%$ relative
to the upper limit of the $V_{SHM}$ range, whereas for the last
listed one (40-50 \%) the corresponding overestimation equals
$\approx 28\%$.

\begin{table}
\caption{\label{Table2} Comparison of predicted volumes of the
hadron source for PHENIX at $\sqrt{s_{NN}}=200$ GeV. Values of
$dV/dy$ are the SHM estimates taken from
Ref.~\cite{Rafelski:2004dp}. For all volumes the unit is cubic
Femtometers. }
\begin{ruledtabular}
\begin{tabular}{ccccc} \hline Centrality [\%] &
$N_{part}$ & $V_{stat.}$ & $V_{SHM} $ & $dV/dy$
\\
\hline 0-5 & 351.4 & 13152 & 8233-12349 & 1920
\\
5-10 & 299.0 & 11279 & 6899-10349 & 1609
\\
10-15 & 253.9 & 9632 & 5694-8542 & 1328
\\
15-20 & 215.3 & 8250 & 4961-7442 & 1157
\\
20-30 & 166.6 & 6391 & 3666-5499 & 855
\\
30-40 & 114.2 & 4381 & 2358-3538 & 550
\\
40-50 & 74.4 & 2812 & 1466-2200 & 342
\\
\hline
\end{tabular}
\end{ruledtabular}
\end{table}

\section {Conclusions}
\label{Conclud}

The single freeze-out model has been applied to estimate
transverse energy and charged particle multiplicity densities as
well as the total multiplicity of charged particles for different
centrality bins of RHIC measurements at $\sqrt{s_{NN}}=130$ and
$200$ GeV. These three variables are independent observables,
which means that they are measured independently of identified
hadron spectroscopy. Since model fits were done to identified
hadron data (particle yield ratios and $p_{T}$ spectra) and the
global variables are calculable in the single freeze-out model, it
was natural to check whether their estimated values agree with the
data. Generally the answer is yes, at least on the qualitative
level. This means that the similar shapes of the functional
dependence on the centrality of the model predictions have been
obtained, only their normalizations differ $\approx 10 \%$ from
the experimental data. In particular, the most surprising result
is the almost ideal independence of the total charged-particle
multiplicity per participating pair on the centrality of the
collision in the range of centralities covered by the PHOBOS data
\cite{Back:2003xk}. The range is $N_{part} \approx 60-360$, that
is from the bin around $50 \%$ centrality up to the most central
one.

Even for $dN_{ch}/d\eta\vert_{mid}$ and $dE_{T}/d\eta\vert_{mid}$
this analysis is nontrivial and does not mean simply a
self-consistency check of various measurements. First, transverse
momentum spectra are measured in \emph{limited ranges}, so very
important low-$p_{T}$ regions are not covered by the data. For
instance, for PHENIX the lowest pion point is at $p_{T}= 0.25$
GeV/c, the lowest kaon one at $p_{T}= 0.45$ GeV/c and the proton
and antiproton data start at $p_{T}= 0.65$ GeV/c
\cite{Adcox:2001mf,Adler:2003cb}. Therefore, to obtain integrated
yields some extrapolations outside the measured ranges are used.
In fact these extrapolations are only analytical fits without any
physical reasoning, but, for instance, contributions from regions
covered by them account for $30 \%$ of the yield for pions, $40
\%$ for kaons and $25 \%$ for protons and antiprotons for RHIC at
$\sqrt{s_{NN}}=130$ GeV \cite{Adcox:2001mf}. On the other hand, a
calorimeter acts very effectively for these species in the
low-$p_{T}$ range, namely pions with $p_{T} \leq 0.35$ GeV/c,
kaons with $p_{T} \leq 0.64$ GeV/c and protons and antiprotons
with $p_{T} \leq 0.94$ GeV/c deposit all their kinetic energy
\cite{Adcox:2001ry}. Second, it is impossible to check the
consistency of the transverse energy data because not all stable
hadron spectra are measured at midrapidity for each collision
case. This mainly concerns neutrons and $K_{L}^{0}$. Since the
good predictions for the transverse energy density at midrapidity
have been obtained (see Figs.~\ref{Fig.7}-\ref{Fig.8}), the
present studies can be understood as an undirect proof that in
these unmeasurable $p_{T}$ regions spectra are also explicable by
means of the thermal source with flow and decays. This has been
also confirmed directly in the PHENIX case by the extension of the
theoretical momentum distributions to the region of the very low
transverse momentum covered by the PHOBOS detector (0.03-0.05 GeV
for pions, 0.09-0.13 GeV for kaons and 0.14-0.21 GeV for protons
and antiprotons), see Fig.~\ref{Fig.4}.

Having compared the present results with the preliminary ones
\cite{Prorok:2004wi}, one can notice that substantial improvement
has been achieved for the PHENIX case at $\sqrt{s_{NN}}=200$ GeV.
In the previous analysis the behavior of the estimates of
$dE_{T}/d\eta\vert_{mid}$ and $dN_{ch}/d\eta\vert_{mid}$ is very
irregular in the region from peripheral to mid-central collisions
and does not give the evidence for any clear dependence on the
centrality. Here the dependence of predictions is exactly the same
as the dependence of the data, only for $dN_{ch}/d\eta\vert_{mid}$
the underestimation of at most $15 \%$ has been obtained. But the
main success of this final analysis is the prediction of the
scaling of the total charged-particle multiplicity with the number
of participants for all considered cases, which failed in the
studies based on the preliminary fits from \cite{Baran:2003nm}.

As the next point, the possible description of the HBT radii
within the single-freeze-out model will be briefly discussed (for
a general review of the HBT interferometry see, \textit{e.g.},
\cite{Baym:1997ce}). This problem needs some comments since the
HBT analysis also gives information about the size of the hadron
source. Studies of the HBT radii in the single-freeze-out model
have been already done for the PHENIX data at $\sqrt{s_{NN}}=130$
GeV \cite{Broniowski:2002wp}. It has turned out that the values of
the estimated radii are about 30 \% smaller then the measured
ones. This discrepancy has been improved by the introduction of
the excluded-volume corrections for the hadron gas, in the way as
postulated in Ref.~\cite{Yen:1997rv}. The corrections give rise to
the appearance of a new common scale factor, denoted as $S^{-3}$,
in the invariant distribution, Eq.~(\ref{Cooper2}). This simply
rescales both geometric parameters $\tau$ and $\rho_{max}$ by the
factor $S$,  \textit{i.e.} after fitting spectra $\tau/S$ and
$\rho_{max}/S$ have the same values as $\tau$ and $\rho_{max}$
obtained for the point-like gas and listed in Table~\ref{Table1}.
The value $S=1.3$ has been found in Ref.~\cite{Broniowski:2002wp}
what implies the increase of $\tau$ and $\rho_{max}$ by 30 \%.
With these new rescaled parameters the predicted values of the HBT
radii agree fairly well with the data \cite{Broniowski:2002wp}. Of
course the detailed analysis of the HBT radii in the context of
the single-freeze-out model for the final RHIC data at
$\sqrt{s_{NN}}=200$ GeV should be performed and this will be the
subject of further investigations.

Finally, it should be notice that the statistical model of the
present analysis corresponds to the chemical equilibrium case of
the more general statistical hadronization model applied in
Ref.~\cite{Rafelski:2004dp} (for a comprehensive review of the
model see \cite{Letessier:2002gp}). In some sense this analysis is
complementary to those studies, because the estimates of
statistical parameters from there have been taken as the input
here and good quality fits have been obtained. Also the
qualitative agreement between independent estimates of the volume
of the hadron source done in both works has been obtained,
\textit{cf.} Table~\ref{Table2}. But as it was shown in
Ref.~\cite{Rafelski:2004dp}, the assumption of the strange quark
non-equilibrium and further also the light quark non-equilibrium
leads to more thorough description of particle yields for all
accessible centralities of PHENIX measurements at
$\sqrt{s_{NN}}=200$ GeV. Therefore such generalization of the
present model should be done and tested with the data and this
will be the subject of the next publication.

To summarize, the single freeze-out version of a statistical
(equilibrium) model fairly well explains the measured spectra of
identified hadrons for various centralities and the centrality
dependence of transverse energy and charged particle multiplicity
pseudo-rapidity densities at mid-rapidity observed at RHIC. The
fact which should be stressed again is that this model predicts
the centrality independence of the total charged-particle
multiplicity. Also predicted values of the total charged-particle
multiplicity per participating pair agree with the measured ones
within $\approx 10 \%$. This is remarkable since geometric
parameters have been fitted to spectra measured at
\textit{midrapidity} but the total charged-particle multiplicity
deals with \textit{the whole rapidity range}.

\begin{acknowledgments}
The author would like to thank Jan Rafelski for very helpful
comments. This work was supported in part by the Polish Committee
for Scientific Research under Contract No. KBN 2 P03B 069 25.
\end{acknowledgments}

\end{document}